\documentclass[twocolumn,amsmath,amssymb,prd]{revtex4}

\def\al{\alpha}
\def\be{\beta}
\def\ga{\gamma}
\def\de{\delta}
\def\ep{\epsilon}

\def\th{\theta}

\def\ka{\kappa}
\def\la{\lambda}

\def\rh{\rho}

\def\ta{\tau}

\def\om{\omega}
\def\Ga{\Gamma}
\def\De{\Delta}

\def\La{\Lambda}

\def\Om{\Omega}

\def\ab{{\al\be}}

\def\vev#1{\langle {#1}\rangle}

\def\lsim{\mathrel{\rlap{\lower4pt\hbox{\hskip1pt$\sim$}}
    \raise1pt\hbox{$<$}}}
\def\gsim{\mathrel{\rlap{\lower4pt\hbox{\hskip1pt$\sim$}}
    \raise1pt\hbox{$>$}}}

\def\etal {{\it et al.}}
\newcommand{\beq}{\begin{equation}}
\newcommand{\eeq}{\end{equation}}
\newcommand{\bea}{\begin{eqnarray}}
\newcommand{\eea}{\end{eqnarray}}
\newcommand{\bse}{\begin{subequations}}
\newcommand{\ese}{\end{subequations}}

\def\sF#1#2{{\textstyle{{#1}\over{#2}}\,}}
\newcommand{\BB}{\big}
\newcommand{\nn}{\nonumber}

\def\to{\rightarrow}

\def\nub{\bar\nu}

\def\aL{(a_L)}
\def\cL{(c_L)}
\def\nub{\bar\nu}

\def\C#1{({\cal C})_{#1}}

\def\aR{(a_R)}
\def\cR{(c_R)}

\usepackage{amsmath}
\usepackage{amsfonts}
\usepackage{amssymb}

\usepackage{float}
\usepackage{graphicx}
\usepackage[usenames,dvipsnames]{color}
\usepackage{wrapfig}

\usepackage{slashed}

\newcommand{\bM}{\begin{pmatrix}}
\newcommand{\eM}{\end{pmatrix}}

\newcommand{\ha}{\sF{1}{2}}

\newcommand{\of}{\text{of}}
\newcommand{\eff}{\text{eff}}

\newcommand{\wT}{\om_\oplus T_\oplus}

\def\A#1#2#3#4{(a^{(#1)}_{#2})_{#3}^{#4}}
\def\C#1#2#3#4{(c^{(#1)}_{#2})_{#3}^{#4}}

\newcommand{\x}{\times}

\begin{document}

\title{Neutrinos as probes of Lorentz invariance}
\author{Jorge S. D\'iaz}
\affiliation{Institute for Theoretical Physics, Karlsruhe Institute of Technology, 76128 Karlsruhe, Germany}
\affiliation{Physics Department, Indiana University, Bloomington, IN 47405, USA}


\begin{abstract}

Neutrinos can be used to search for deviations from exact Lorentz invariance.
The worldwide experimental program in neutrino physics makes these particles a remarkable tool to search for a variety of signals that could reveal minute relativity violations.
This paper reviews the generic experimental signatures of the breakdown of Lorentz symmetry in the neutrino sector.

\end{abstract}

\maketitle

\section{Introduction}

In 1930, Pauli postulated the neutrino as a {\it desperate remedy} to save one of the sacred principles of physics: the conservation of energy, which appeared to be violated in beta decays \cite{Pauli}.
Today, neutrinos remain as some of the less understood particles of the Standard Model (SM) and their mysteries are still fascinating.
Their ghostly nature make them barely interact with matter and their interferometric behavior make them oscillate between different flavors.
Neutrino oscillations have led to the remarkable conclusion of massive neutrinos, presenting an established evidence of physics beyond the SM.

In the search for new physics, different candidate theories for quantum gravity involve mechanisms that could trigger the breakdown of one of the most fundamental symmetries in modern physics: Lorentz invariance.
In the theoretical front, Lorentz-violating descriptions of neutrino behavior have shown that these fundamental particles can serve as powerful probes of new physics. 
Experimentally, neutrino oscillations have been used to perform several searches for Lorentz violation.
The development of techniques to perform systematic searches for Lorentz violation in many other experimental setups show a rich phenemenology to be studied, with a large variety of experimental effects that remain unexplored.

This paper summarizes the main experimental signatures of deviations from exact Lorentz invariance in the neutrino sector.
The experimental searches for Lorentz violation performed in recent years are presented and future tests of Lorentz symmetry are discussed, ranging from precision measurements of beta decay and double beta decay at low energies to the high energy of astrophysical neutrinos and oscillation experiments using accelerator, atmospheric, and reactor neutrinos.

\section{Lorentz invariance violation}

Deviations from exact Lorentz symmetry have been shown to be possible at very high energies in candidate descriptions of gravity at the quantum level.
For instance, mechanisms for the spontaneous breaking of this fundamental symmetry have been identified in string-theory scenarios \cite{SBS_LV1}. 
Interactions could generate Lorentz-violating terms if a tensor field acquires a nonzero vacuum expectation value $\vev{T_{\al\be\ga\cdots}}=t_{\al\be\ga\cdots}\neq0$, which acts as a background field.
In the same fashion as the nonzero vacuum expectation value of the dynamical Higgs field generates mass terms for other fields via interactions, background tensor fields that couple to conventional particles in the SM will generate new terms that break Lorentz invariance.
These new terms are Lorentz scalars under coordinate transformations; in fact, the spacetime indices of the background field are all contracted with the indices of the SM operator in the form $t_{\al\be\ga\cdots}\mathcal{O}^{\al\be\ga\cdots}$. 
This structure guarantees that there is no privileged reference frame because the theory is observer invariant.
For instance, consider the case of Lorentz violation generated by a 2-tensor $t_{\al\be}$, which will be coupled to some SM operator with the same number of spacetime indices in the form $\mathcal{L}=t_\ab\mathcal{O}^\ab$.
Under a coordinate transformation both the operator and the tensor background field transform to a new set of coordinates as
\bea\label{LT}
\mathcal{O}^{\ab} &\to& \mathcal{O}^{\al'\be'} = 
\La^{\al'}_{\phantom{\al'}\al}\La^{\be'}_{\phantom{\be'}\be} \; \mathcal{O}^{\ab},
\nn\\
t_{\ab} &\to& \, t_{\al'\be'} \;= (\La^{-1})^{\la}_{\phantom{\la}\al'}(\La^{-1})^{\rh}_{\phantom{\rh}\be'}\; t_{\la\rh},
\eea
so that the terms in the lagrangian $\mathcal{L}=t\cdot\mathcal{O}$ remain invariant $\mathcal{L}'=t'\cdot\mathcal{O}'=t\cdot\mathcal{O}=\mathcal{L}$.
The same construction can be applied for a general tensor.

On the other hand, when a Lorentz transformation is performed over the physical system, this is when the experimental apparatus is rotated or boosted rather than the coordinates used to describe it, then the SM operator transforms as shown in \eqref{LT} but any background field remains unchanged $\mathcal{L}'=t\cdot\mathcal{O}'\neq\mathcal{L}$.
This so-called particle Lorentz transformation changes the coupling between the background fields and the SM operators, resulting in a physically observable anisotropy of spacetime; this is a violation of Lorentz invariance \cite{SME1a}.

Phenomenological approaches to parametrize and experimentally search for particular types of Lorentz violation have been considered since several decades \cite{LNI1,LNI2,LNIexp1,LNIexp2}.
However, effective field theory can be used to incorporate generic operators that break Lorentz invariance for all the particles in the SM.
This general framework is known as the Standard-Model Extension (SME) \cite{SME1a,SME1b,SME2}, whose action includes general coordinate-invariant terms by contracting operators of conventional fields with controlling coefficients for Lorentz violation and reduces to the SM if all these coefficients vanish.
Gravity can also be incorporated by writing the SM in a curved background \cite{SME2}.
The development of the SME has led to a worldwide experimental program searching for violations of Lorentz invariance, whose results are summarized in Ref. \cite{tables}. 

Flat spacetime is considered for experiments in particle physics, in which case the coefficients for Lorentz violation that act as background fields can be chosen to be constant and uniform, which guarantees conservation of energy and linear momentum.
In this limit, these coefficients represent the vacuum expectation value of the tensor fields of the underlying theory.
Excitations of these fields lead to a rich phenomenology, for instance Nambu-Goldstone modes could play fundamental roles when gravity is included, such as the graviton, the photon in Einstein-Maxwell theory \cite{NGmodes2,NGmodes3,NGmodes5,NGmodes6}, spin-dependent \cite{spin_dep} and spin-independent \cite{spin_indep} forces.

It should be noted that a subset of operators in the SME also break CPT symmetry.
In fact, all the Lorentz-violating terms in the action involving operators with an odd number of spacetime indices are odd under a CPT transformations.
In realistic field theories, CPT violation always appears with Lorentz violation \cite{CPTv}.
Nonetheless, alternative approaches exist in which CPT violation is implemented with and without Lorentz invariance \cite{CPT01,CPT02,CPT03,CPT04,CPT06,CPT08,CPT10,CPT11}.

\section{Neutrinos}
\label{Sec_nus}

The general description of three left-handed neutrinos and three right-handed antineutrinos in the presence of Lorentz-violating background fields is given by a $6\x6$  effective Hamiltonian of the form \cite{LVnu}
\beq\label{H_SME}
H=\bM (h_0)_{ab} & 0 \\ 0 & (h_0)_{ab}^* \eM
+\bM \de h_{ab} & \de h_{a\bar b} \\ \de h_{a\bar b}^* & \de h_{\bar a\bar b} \eM,
\eeq
where the indices indicate the flavors of active neutrinos $a,b=e,\mu,\ta$ and antineutrinos $\bar a,\bar b=\bar e,\bar\mu,\bar\ta$.
The Lorentz-preserving component is explicitly given by
\beq\label{h0}
(h_0)_{ab} = |\pmb{p}|\de_{ab} +\frac{m^2_{ab}}{2|\pmb{p}|}, 
\eeq
where at leading order the neutrino momentum is given by the energy $|\pmb{p}|\approx E$ and the mass-squared matrix is commonly written in terms of the Pontecorvo-Maki-Nakagawa-Sakata (PMNS) matrix \cite{PMNS1,PMNS2} as $m^2=U_\text{PMNS}(m^2_D)U_\text{PMNS}^\dag$, with $m^2_D=\text{diag}(m_1^2,m_2^2,m_3^2)$.

The Lorentz-violating block describing neutrinos in the Hamiltonian \eqref{H_SME} is given by
\beq\label{h_SME}
\de h_{ab} = \aL^\al_{ab} \hat p_\al + \cL^\ab_{ab}\hat p_\al\hat p_\be|\pmb{p}|.
\eeq
The components of the $3\times3$ complex matrices $\aL^\al_{ab}$ and $\cL^\ab_{ab}$ are called coefficients for CPT-odd and CPT-even Lorentz violation, respectively. 
The spacetime indices $\al,\be$ encode the nature of the broken symmetry; for instance, isotropic (direction-independent) Lorentz violation appears when only the time components of the coefficients are nonzero; while space anisotropy appears when any of the other components is nonzero, generating direction-dependent effects in the neutrino behavior.
The breakdown of invariance under rotations is evident due to the presence of the four-vector $\hat p^\al=(1;\hat p)$ that depends on the neutrino direction of propagation $\hat p$.

The block Hamiltonian describing right-handed antineutrinos is obtained as the CP conjugate of the neutrino Hamiltonian $\de h_{\bar a\bar b}=\text{CP}(\de h_{ab})$, which has the same form as the neutrino Hamiltonian \eqref{h_SME} with $\aR^\al_{\bar a\bar b}=-\aL^{\al*}_{ab}$, $\cR^\ab_{\bar a\bar b}=\cL^{\ab*}_{ab}$.
Given the structure of these coefficients in flavor space, it is expected that they will affect neutrino mixing and oscillations.
Notice, however, that there exist coefficients that modify the three flavors in the same way, producing no effects on neutrino oscillations because they are proportional to the identity in flavor space.
These {\it oscillation-free} coefficients and their observable effects are discussed in Sec. \ref{Sec_nu_velocity} and \ref{Sec_WDecays}.

Signals of the breakdown of Lorentz invariance correspond to the anisotropy of spacetime due to preferred directions set by the coefficients for Lorentz violation that act as fixed background fields.
Taking advantage of the coupling of these background fields with the neutrino direction of propagation $\hat p$, we can search for violations of Lorentz invariance by making measurements with neutrino beams with different orientations, which would reveal the presence of the SME coefficients $\aL^\al_{ab}$ and $\cL^\ab_{ab}$.
For Earth-based experiments, detectors and source rotate with a well-defined angular frequency $\om_\oplus\simeq2\pi/(\text{23 h 56 min})$ due to Earth's rotation, which makes the neutrino direction vary with respect to the fixed background fields. This time dependence will explicitly appear in the relevant observable quantities and it can be parametrized as harmonics of the sidereal angle $\wT$.
Due to the invariance of the theory under coordinate transformations, there is no preferred reference frame to make the measurements.
In order to establish a consistent and systematic search for Lorentz-violating effects, experimental results are conventionally reported in the Sun-centered equatorial frame described in Refs. \cite{tables,SunFrame1}.
In this frame, the sidereal variation of the coupling between the neutrino direction $\hat p$ and the background fields that break Lorentz symmetry can be explicitly written in the form
\bea\label{dh(wT)}
\de h_{ab}&=& (\mathcal{C})_{ab}+(\mathcal{A}_s)_{ab}\sin\wT+(\mathcal{A}_c)_{ab}\cos\wT \nn\\
&&+(\mathcal{B}_s)_{ab}\sin2\wT+(\mathcal{B}_c)_{ab}\cos2\wT,
\eea 
where the amplitude of each sidereal harmonic is a function of the coefficients for Lorentz violation and experimental parameters including neutrino energy, location of the experiment, and relative orientation between source and detector \cite{KM_SB,DKM}.

The coefficients $\aL^\al_{ab}$ and $\cL^\ab_{ab}$ arise from operators of dimension three and four, respectively.
Operators of arbitrary dimension $d$ can be incorporated in the theory, in which case the coefficients for Lorentz violation in the Hamiltonian \eqref{h_SME} appear as momentum-dependent quantities of the form \cite{KM2012,KM2013}
\bea
\aL^\al_{ab} &\to& (\hat a_L)_{ab}^{\al\la_1\ldots\la_{d-3}} p_{\la_1}\cdots p_{\la_{d-3}},\quad d\text{ odd},
\nn\\
\cL^\ab_{ab} &\to& (\hat c_L)_{ab}^{\ab\la_1\ldots\la_{d-3}} p_{\la_1}\cdots p_{\la_{d-3}},\quad\!\! d\text{ even}.
\eea
The extra derivatives in the Lagrangian appear in the neutrino Hamiltonian as higher powers of the neutrino energy.
Although the conventional massive-neutrino description of oscillations accommodates all the established experimental results, non-negative powers of the neutrino energy could help to elegantly solve some anomalous results obtained in recent years in beam experiments \cite{LSND_anomaly,MB_anomaly1,MB_anomaly2}.
In fact, interesting attempts to describe the global data using the SME have led to the construction of alternative models for neutrino oscillations that can accommodate the results reported by different experiments \cite{LVmodels1,LVmodels2,LVmodels3,LVmodels4,LVmodels5,LVmodels6,LVmodels7}.

In the following sections we discuss the observable signatures of these coefficients for Lorentz violation in different types of neutrino experiments including oscillations, neutrino velocity, and beta decays.

\section{Neutrino oscillations}

The interferometric nature of neutrino oscillations has been widely identified as an ideal experimental setup to search for new physics in the form of deviations from the conventional description of neutrinos.
The mixing and oscillation between neutrino flavors occurs in general due to off-diagonal entries in the neutrino Hamiltonian leading to eigenstates with different energy.

\subsection{Oscillation of neutrinos and antineutrinos}

Neutrino oscillations are well described by a model of three massive neutrinos, which depends on two mass-squared differences $\De m^2_{21}$ and $\De m^2_{31}$ controlling the oscillation lengths, and three mixing angles $\th_{12},\th_{13},\th_{23}$ that govern the amplitude of the oscillation \cite{PDG2012}.
According to this massive-neutrino model, the oscillation probabilities are proportional to the factor $\sin^2(\De m^2_{ij}L/4E)$.
Tests of Lorentz invariance using neutrino oscillations can be performed by searching for deviations from the conventional  behavior.
For some experiments studying neutrinos over a large range of energies and baselines, such as Super-Kamiokande \cite{LV_SK}, an exact treatment of the diagonalization of the Hamiltonian is necessary \cite{ExactDiag}.
In most cases, the experimental features allow the implementation of approximation methods for determining the oscillation probabilities, as discussed in the following sections.

\subsubsection{Short-baseline approximation}

According to the massive-neutrino model, for experiments using neutrinos of energy $E$ and baseline $L$ that satisfy $\De m^2_{ij}L/4E\ll\pi/2$, the oscillation phase would be too small to impact the neutrino propagation and no neutrino oscillations should be observed.
The effects of the mass terms in the conventional Hamiltonian $h_0$ \eqref{h0} become negligible and the effective Hamiltonian can be approximated by $h\approx\de h$.
Direct calculation of the oscillation probabilities shows that in this approximation we can write the appearance probability as \cite{KM_SB}
\beq\label{P_SB}
P_{\nu_b\to\nu_a}\simeq L^2|\de h_{ab}|^2,\quad a\neq b,
\eeq
with a similar expression for antineutrinos using $\de h_{\bar a\bar b}$ instead.
The sidereal decomposition of the Hamiltonian \eqref{dh(wT)} can be used to show that the probability will also exhibit sidereal variations, one of the key signatures of Lorentz violation.
The Hamiltonian contains also a time-independent component $(\mathcal{C})_{ab}$, which can lead to both isotropic and direction-dependent effects. 
In all cases, the energy dependence is different with respect to the conventional case, so spectral studies can be used to study these particular coefficients.

Experimental studies using the probability \eqref{P_SB} have been performed by Double Chooz \cite{LV_DC}, IceCube \cite{LV_IceCube}, LSND \cite{LV_LSND}, MiniBooNE \cite{LV_MiniBooNE1,LV_MiniBooNE2}, and MINOS using its near detector \cite{LV_MINOS_ND1,LV_MINOS_ND2}.
The absence of a positive signal in all these experiments has been used to set tight constraints on several coefficients for Lorentz violation, which are summarized in Ref. \cite{tables}.
It is important to emphasize that since all these searches use different oscillation channels, they are complementary, accessing similar coefficients but with different flavor indices.

\subsubsection{Perturbative approximation}
\label{Sec_perturb}

Since the phase of the oscillation is given by $\De m^2_{ij}L/4E$, the oscillation probability can be enhanced by placing a detector at a distance $L=2\pi E/\De m^2_{ij}$ from the neutrino source.
For experiments satisfying this condition, mass-driven oscillations $h_0$ dominate in the Hamiltonian and we can consider the effects of Lorentz violation $\de h$ as a small perturbation \cite{DKM}.
In this case, the oscillation probability appears as a power series of the form
\beq
P_{\nu_b\to\nu_a} = P_{\nu_b\to\nu_a}^{(0)} + P_{\nu_b\to\nu_a}^{(1)} + \ldots,
\eeq
where $P_{\nu_b\to\nu_a}^{(0)}$ is the conventional probability given by the massive-neutrino model and the following terms are given in powers of the coefficients for Lorentz violation, whose explicit form is given in Ref. \cite{DKM}.
Once again, the sidereal variation of the oscillation probability appears as a key signal to search by experiments.
For instance, the leading-order term $P_{\nu_b\to\nu_a}^{(1)}$ can be generically written as
\bea\label{P1}
&&\frac{P_{\nu_b\to\nu_a}^{(1)}}{2L} = (P_{\mathcal{C}}^{(1)})_{ab} \nn\\
&&\quad\quad
+(P_{\mathcal{A}_s}^{(1)})_{ab}\sin{\wT}%
+(P_{\mathcal{A}_c}^{(1)})_{ab}\cos{\wT} \nn\\
&&\quad\quad
+(P_{\mathcal{B}_s}^{(1)})_{ab}\sin{2\wT}
+(P_{\mathcal{B}_c}^{(1)})_{ab}\cos{2\wT}.
\eea
This is the dominating probability for neutrino mixing as well as antineutrino oscillations.

\begin{figure}
\centering
\includegraphics[width=0.5\textwidth]{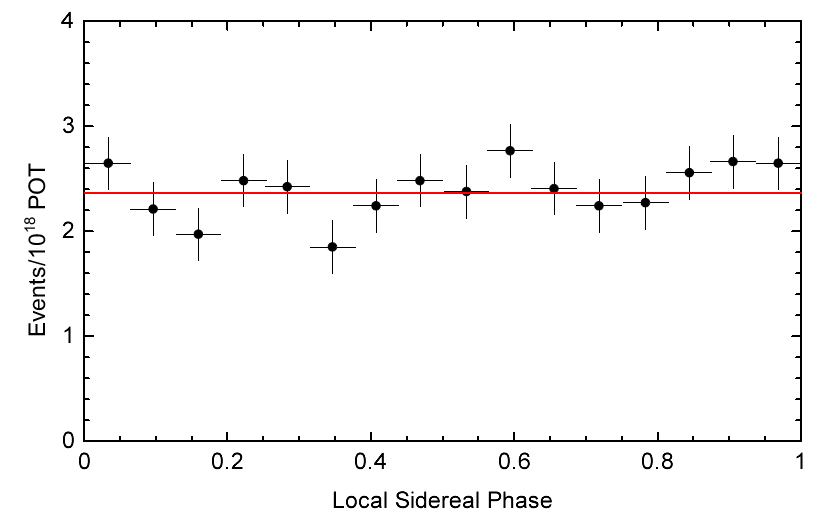}
\caption{Number of events normalized by protons on target (POT) in the MINOS far detector as a function of the sidereal phase. 
The flat distribution of events is interpreted as the absence of sidereal variations in the oscillation probability.
Figure adapted from Ref. \cite{LV_MINOS_FD}.} 
\label{Figure_MINOS}
\end{figure}

Since the first-order correction \eqref{P1} to the oscillation probability arises from the interference between the conventional and the Lorentz-violating effects, the sensitivity to the coefficients in $P_{\nu_b\to\nu_a}^{(1)}$ is greater than in the short-baseline approximation presented in the previous section.
Figure \ref{Figure_MINOS} shows part of the study performed by the MINOS experiment, which used the expression \eqref{P1} to search for sidereal variations in the event rate measured at the far detector.
The sensitivity to different coefficients was improved by a factor 20-510 compared to the previous constraints using the near detector \cite{LV_MINOS_FD}.

The mixing between neutrinos and antineutrinos is also possible due to the block $\de h_{a\bar b}$ in the hamiltonian \eqref{H_SME}, which is discussed in the following section.

\subsection{Neutrino-antineutrino mixing}

The off-diagonal block $\de h_{a\bar b}$ in the Hamiltonian \eqref{H_SME} can produce the mixing between neutrinos and antineutrinos.
This $3\x3$ matrix is given by \cite{LVnu}
\beq\label{h(g,H)}
\de h_{a\bar b} =
i\sqrt2 (\ep_+)_\al\tilde{H}^\al_{a\bar b}-i\sqrt2(\ep_+)_\al\hat p_\be\,\tilde g^\ab_{a\bar b} |\pmb{p}|,
\eeq
where the complex 4-vector $(\ep_+)_\alpha$ is the neutrino polarization that can be directly written in terms of the location of the experiment and the orientation of the neutrino beam \cite{DKM}.
Two sets of coefficients for Lorentz violation denoted by $\tilde{H}^\al_{a\bar b}$ and $\tilde g^\ab_{a\bar b}$ control CPT-even and CPT-odd effects, respectively.
This Hamiltonian can also be decomposed in the form \eqref{dh(wT)}, with harmonic amplitudes given in terms of the coefficients $\tilde{H}^\al_{a\bar b}$ and $\tilde g^\ab_{a\bar b}$ \cite{DKM}.

Contrary to the neutrino Hamiltonian \eqref{h_SME}, the neutrino-antineutrino block always appears with direction-dependent effects, for this reason the search of sidereal variations is an ideal setup to search for these coefficients.
Following the perturbative description presented in Sec. \ref{Sec_perturb}, it has been shown that at first order the oscillation probability vanishes, in other words, neutrino-antineutrino oscillations appear as a second order effect \cite{DKM}.
For this reason, the second-order probability $P_{\nu_a\to\nub_b}^{(2)}$ can be decomposed in the form \eqref{P1}, although involving up to fourth harmonics.

\begin{figure}
\centering
\includegraphics[width=0.46\textwidth]{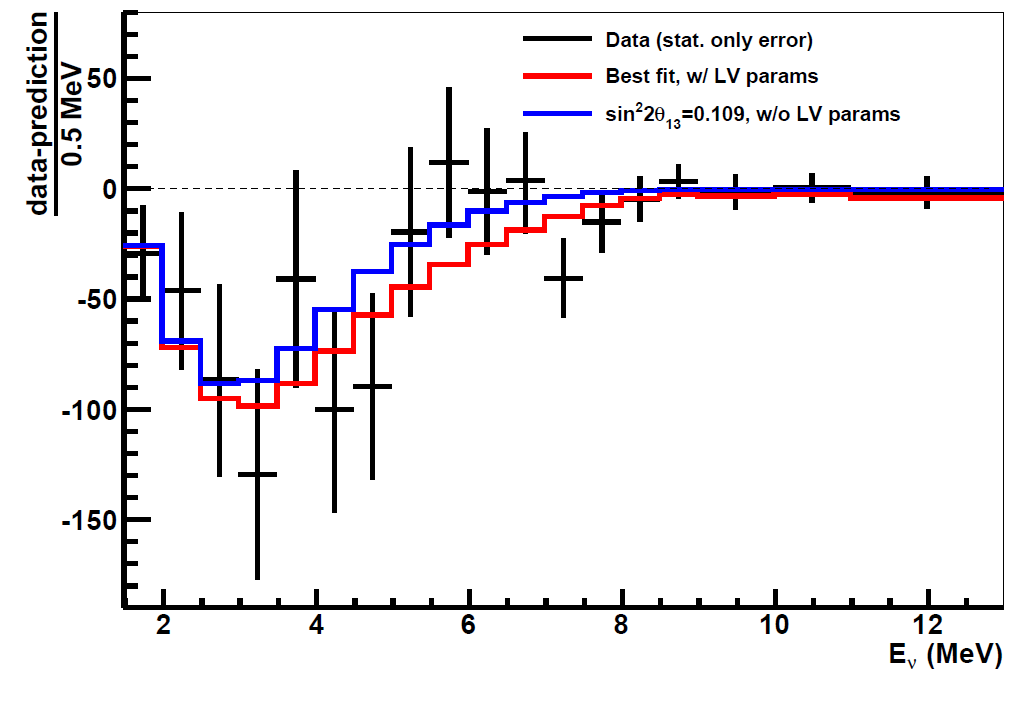}
\caption{Fit to the disappearance of reactor antineutrinos in the form $\nub_e\to\nu_e$ due to neutrino-antineutrino mixing (red line) and the conventional oscillation in the absence of Lorentz violation (blue line) in the Double Chooz experiment. Figure adapted from Ref. \cite{LV_DC2}.} 
\label{Figure_DC}
\end{figure}

The possible oscillation of neutrinos into antineutrinos modifies, for instance, the survival probability of muon neutrinos in a beam experiment because now some $\nu_\mu$ could disappear into antineutrino states.
A systematic search of the 66 coefficients $\tilde{H}^\al_{a\bar b}$ and $\tilde g^\ab_{a\bar b}$ producing sidereal variations was performed using data from the MINOS experiment \cite{RebelMufson}.
The remaining 15 coefficients producing time-independent effects could only be explored by a spectral study in a disappearance experiment.
Figure \ref{Figure_DC} shows a fit to the data from the Double Chooz experiment, searching for the spectral modification that could arise in the disappearance of electron antineutrinos \cite{LV_DC2}.

A total of 81 coefficients $\tilde{H}^\al_{a\bar b}$ and $\tilde g^\ab_{a\bar b}$ have been tightly constrained by these two experimental searches, whose results are summarized in Ref. \cite{tables}.

\section{Neutrino kinematics}
\label{Sec_nu_velocity}

Oscillations are very sensitive to unconventional effects producing neutrino mixing due to their interferometric nature.
There are, however, terms in the Hamiltonian \eqref{H_SME} that are unobservable in oscillations.
Neutrino oscillations only allow us to measure energy differences between different neutrino states, for this reason the absolute scale of neutrino masses cannot be determined from oscillations.
Similarly, some coefficients for Lorentz violation modify the energy of all flavors in the same way producing no effects in oscillations.
Neglecting mixing effects results in a decoupling of the three-flavor system into three copies of a single state.
One of the observable effects of these {\it oscillation-free} coefficients is the modification of the neutrino velocity, which 
produces measurable effects in the neutrino time of flight.
Moreover, as a consequence of the unconventional dispersion relations, the neutrino phase space and energy-conservation condition relevant for decay processes are modified as well.

\subsection{Neutrino velocity}

The neutrino velocity can be obtained from the Hamiltonian \eqref{H_SME}.
For completeness, operators of arbitrary dimension $d$ can be incorporated, in which case the neutrino velocity takes the form \cite{KM2012}
\bea\label{v}
v_\nu &=& 1 - \frac{|m|^2}{2|\pmb{p}|^2} +
\sum_{djm} (d-3)|\pmb{p}|^{d-4}\,e^{im\wT}\,_0\mathcal{N}_{jm} \nn\\
&&\qquad\qquad\qquad\quad
\times\BB(\A{d}{\of}{jm}{}-\C{d}{\of}{jm}{}\BB),
\eea
where the factor $|m|^2$ is a real mass parameter that does not participate in oscillations, and the Lorentz-violating component has been written in spherical form. The index $d$ denotes the effective dimension of the operator and the pair $jm$ corresponds to angular momentum indices that label the rotational properties of the oscillation-free spherical coefficients $\A{d}{\of}{jm}{}$ and $\C{d}{\of}{jm}{}$, controlling CPT-odd (for odd $d$) and CPT-even (for even $d$) effects, respectively.
These spherical coefficients can be identified with coefficients in cartesian coordinates used in the previous sections \cite{KM2012}.
The expression for the neutrino velocity \eqref{v} has been written in the Sun-centered frame \cite{tables,SunFrame1}, where all the directional information is contained in the angular factors $_0\mathcal{N}_{jm}$ and the sidereal time dependence appears as harmonics functions controlled by the index $m$.

The neutrino velocity \eqref{v} exhibits a rich phenomenology in the form of many physical effects than can affect neutrino propagation if deviations from Lorentz symmetry are present.
Depending on the dimension $d$ of the operator in the theory, neutrino velocity can be energy dependent; for $j\neq0$, anisotropic effects appear and the velocity becomes a function of the direction of propagation; for $m\neq0$ time dependence arises, in which case the neutrino velocity varies with sidereal time $T_\oplus$; and for odd $d$, CPT violation makes neutrinos and antineutrinos move at different speed.

Beam experiments are suitable setups to compare the speed of neutrinos with respect to the speed of photons.
From the neutrino velocity \eqref{v}, we clearly find that the mass term makes neutrinos travel slower than light, whereas the coefficients for Lorentz violation can generate subluminal or superluminal velocities depending on the sign of each coefficient.
Different beam experiments have measured the time for neutrinos to travel a distance $L$ \cite{FNAL76,FNAL79,MINOS_v,OPERA,ICARUS,Borexino,LVD}, which will experience a delay with respect to photons given by
\beq
\De t \approx L(1-v_\nu),
\eeq
which can be used to set limits in the oscillation-free coefficients for Lorentz violation that modify the neutrino velocity in \eqref{v}.
Since the minute effects of Lorentz violation can be enhanced by neutrinos travelling a long distance, a precise constraint on the isotropic dimension-four coefficient was obtained using the few antineutrino events from the supernova SN1987A \cite{Longo_SN}.

For the particular case of Lorentz invariance violation generated by a dimension-four operator, the modification to the neutrino velocity \eqref{v} is simply a constant factor.
For operators of dimension $d\geq5$, low- and high-energy neutrinos will move at different velocity.
If a burst of neutrinos of different energies are created at the same time, this velocity difference will generate an spread of neutrinos, observable as a delay between high- and low-energy neutrinos at the detector \cite{KM2012}.
A similar effect has been widely studied for Lorentz-violating photons \cite{Photon_disp1,Photon_disp2,Photon_disp3,Photon_disp4}.

\subsection{Threshold effects}

The modified dispersion relations that neutrinos satisfy in the presence of Lorentz violation alter the energy-momentum conservation relation, which plays an important role in meson-decay processes of the form $M^+\to l^++\nu_l$.
It can be shown that above some threshold energy $E_\text{th}$ these relations can completely block the phase space available for the decay \cite{KM2012,threshold1,ChodosTh,AKTh,threshold2,threshold3,threshold4,threshold5,threshold6,threshold8}.
The observation of atmospheric and accelerator neutrinos $\nu_l$ with energy $E_0$ produced by the decay of a meson of mass $M_M$, implies that $E_\text{th}>E_0$, which can be used to write the condition \cite{KM2012}
\beq\label{Threshold}
\sum_{djm} E_0^{d-2}\,Y_{jm}(\hat{\pmb{p}}) \BB[\pm\A{d}{\of}{jm}{}-\C{d}{\of}{jm}{}\BB] \lesssim
\sF{1}{2}(M_M-m_{l^\pm})^2,
\eeq
where the $+$ ($-$) sign is for neutrinos (antineutrinos) and $m_{l^\pm}$ is the mass of the accompanying charged lepton.
This formula has been used in Ref. \cite{KM2012} to constrain several coefficients for Lorentz violation, including many associated to nonrenormalizable operators.

The sensitivity to the effects of Lorentz violation increases with the energy of neutrinos observed as well as the distance that they travel.
The observation of very-high-energy neutrinos reported by the IceCube collaboration \cite{IceCube_PeVnus,IceCube_28nus} offers a great sensitivity to the effects described in this section.
The small number of neutrinos observed with energies at the PeV level \cite{IceCube_PeVnus} allows the study of isotropic effects ($j=0$); nevertheless, a full study of direction-dependent effects would require several events spread in the sky.
Although the IceCube results suggest an astrophysical origin for these energetic neutrinos, tight constraints on different coefficients for Lorentz violation can be obtained even in the conservative interpretation of these neutrino events having atmospheric origin \cite{VHEnus}.
The observation of PeV neutrinos created by the decay of heavy mesons in the upper atmosphere has been used to implement the threshold condition \eqref{Threshold}, leading to sensitive limits in several isotropic coefficients of dimension $d=4,6,8$ and $10$ \cite{VHEnus}.

\subsection{\v Cerenkov radiation}

In same way as some processes can be forbidden above certain energies, the effects of Lorentz violation can also open particular decay channels that would be otherwise forbidden.
In particular, coefficients leading to $v_\nu>1$ in \eqref{v} can produce \v Cerenkov emission of one or more particles 
\cite{KM2012,ColemanGlashow,Cerenkov1,Cerenkov2,Cerenkov3,Cerenkov4,Cerenkov5,Cerenkov6,Cerenkov7,Cerenkov8,Cerenkov9,
Cerenkov11}.
\v Cerenkov radiation makes neutrinos lose energy, which distorts the spectrum in long-baseline experiments using accelerator and atmospheric neutrinos.
This feature provides another method to search for Lorentz violation.
The observation of high-energy neutrinos after propagating a distance $L$ sets a lower value for the characteristic distortion distance $D(E)=-E/(dE/dx)$ in the form $L<D(E)$.
The determination of the characteristic distance for the spectral distortion caused by the isotropic Lorentz-violating operator of dimension four $\C{4}{\of}{00}{}$ is described in Refs. \cite{ColemanGlashow,Cerenkov1,Cerenkov2,Cerenkov3,Cerenkov4,Cerenkov5,Cerenkov6,Cerenkov7,Cerenkov8,Cerenkov9,
Cerenkov11}.
The general calculation including direction-dependent effects for operators of arbitrary dimension can be found in Ref. \cite{KM2012}.

\begin{figure}
\centering
\includegraphics[width=0.35\textwidth]{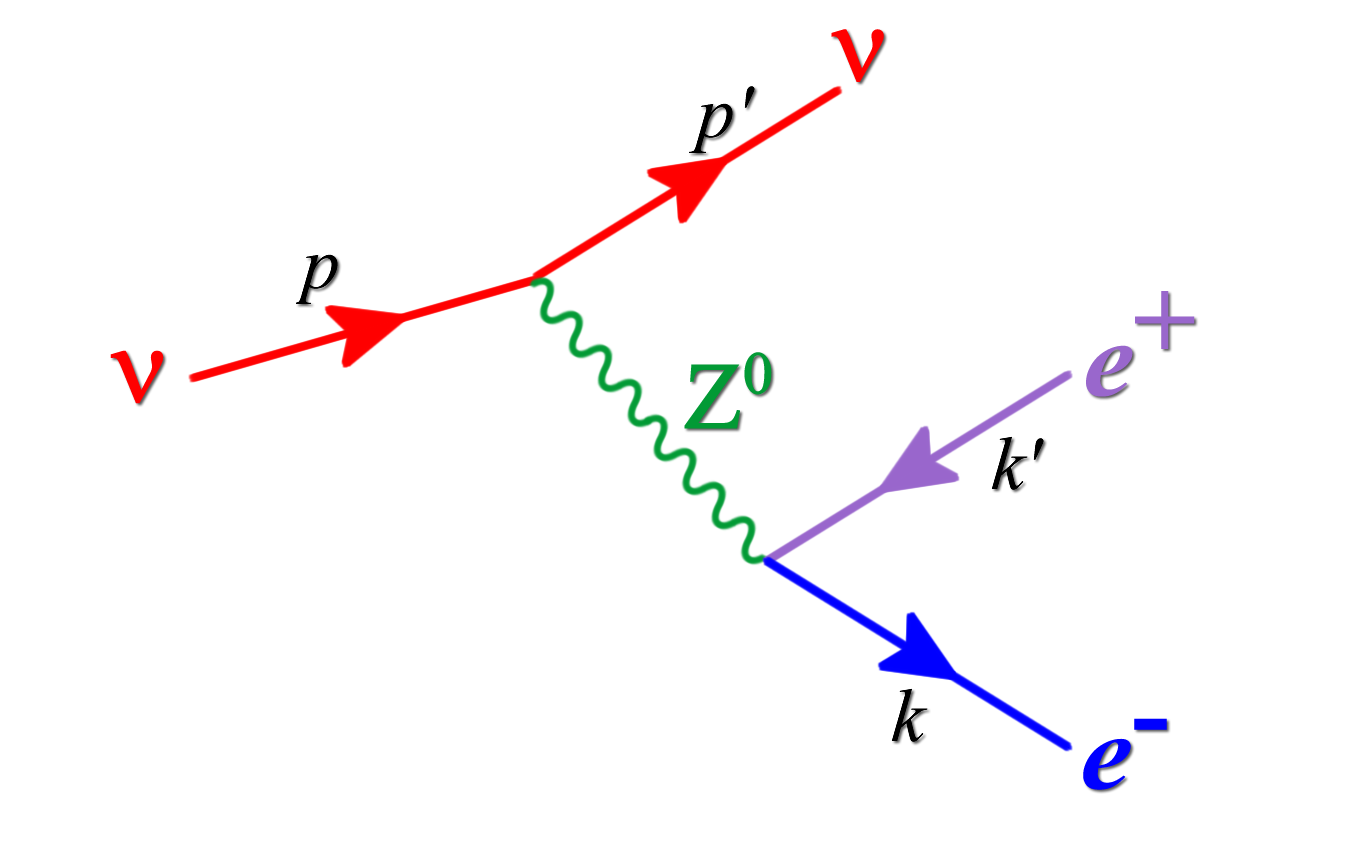}
\caption{Electron-positron pair emission as neutrino \v Cerenkov radiation.} 
\label{Figure_Cerenkov}
\end{figure}

Using the PeV neutrinos observed by IceCube \cite{IceCube_PeVnus}, the limits obtained using threshold conditions can be improved by one order of magnitude by determining spectral distortion produced by \v Cerenkov radiation \cite{VHEnus}.
For instance, the emission of electron-positron pairs in the form $\nu\to\nu+e^-+e^+$ is characterized by a rate of energy loss given by \cite{KM2012,VHEnus}
\beq
\frac{dE}{dx}= -
\frac{C}{8}\!\int\!\!\frac{\ka^0\,\pmb{\ka}'^2}{(\ka^2-M_Z^2)^2}\frac{\partial|\pmb{\ka}'|}{\partial\ka_0}
\frac{q\cdot k\;q'\!\cdot k'}{q_0k_0q_0'k_0'}\,d^3p'\,d\Om_{\ka'},
\eeq
where $C$ is a constant, the auxiliary 4-vectors $\ka=k+k'$ and $\ka'=k-k'$ have been defined in terms of the momentum of the electron and the positron, and $q/q_0=(1,\hat{\pmb{p}})$, $q'/q_0'=(1,\hat{\pmb{p}}')$, following the conventions in Fig. \ref{Figure_Cerenkov}.
Several orders of magnitude in sensitivity can be gained when using an astrophysical interpretation for the PeV neutrinos in IceCube.
After travelling astrophysical distances these neutrinos would rapidly fall below the threshold energy for \v Cerenkov emission.
The observation of these neutrinos with PeV energies implies that this threshold energy lies above 1 PeV, leading to stringent limits on isotropic Lorentz violation of dimension $d=4,6,8$ and $10$ \cite{VHEnus}.
Similar studies for the case $d=4$ can be found in Refs. \cite{Borriello2013,Stecker2013}.

Direction-dependent effects using high-energy neutrinos require several events.
The recent observation of 26 new energetic events in IceCube \cite{IceCube_28nus} distributed in the sky allows the search of space anisotropy for operators of dimension $d=4,6$ \cite{VHEnus}.
The simultaneous study of several coefficients producing direction-dependent effects allows two-sided bounds, more restrictive than the very particular case of isotropic Lorentz violation considering superluminal velocity that allows one-sided limits only.
In the future, the observation of more events should allow a detailed study of operators of higher dimension.

\section{Beta decay}
\label{Sec_WDecays}

The interferometric nature of neutrino oscillations make them an ideal type of experiment to search for minute deviations from exact Lorentz symmetry.
Nonetheless, the effects that modify the kinematics of all neutrino flavors in the same manner are unobservable in oscillation experiments, which makes the studies described in Sec. \ref{Sec_nu_velocity} an important complement to oscillation searches.
The enhancement of Lorentz-violating effects with the neutrino energy makes also the study of neutrino velocity and \v Cerenkov radiation a sensitive probe of Lorentz invariance with high-energy neutrinos.
Nevertheless, it has been shown that low-energy experiments can also play a key role in the study of Lorentz invariance.
In particular, signals of oscillation-free operators of dimension three $\A{3}{\of}{jm}{}$ are not only unobservable in oscillations but also produce no effects in the neutrino velocity \eqref{v}.

The experimental signatures of the coefficients associated to these so-called {\it countershaded} operators \cite{spin_indep,DKL,KT} motivate the study of weak decays.
The effects of these operators are unaffected by the neutrino energy, giving low-energy experiments a competitive sensitivity to signals of Lorentz violation.
It is important to emphasize that Lorentz-violating effects appear as kinematical effects modifying the neutrino phase space; nevertheless, modifications of the spinor solutions must also be taken into account.
Beta decay in the context of Lorentz violation in sectors other than neutrinos have recently been studied theoretically \cite{Noordmans,Altschul} and experimentally \cite{Muller}.

\subsection{Tritium decay}

The absolute mass scale of neutrinos cannot be studied in oscillation experiments, which only offer access to mass-squared differences.
The direct measurement of neutrino masses can be made by searching for a distortion of the electron energy spectrum in tritium decay.
The measurement of beta electrons near the endpoint of the spectrum 
\beq\label{BetaSpectrum}
\frac{d\Ga}{dT} = C\big[(\De T)^2-\ha m_\nu^2\BB],
\eeq
allows the study of the effective mass $m^2_\nu$ of electron antineutrinos, where $C$ is approximately constant and $\De T=T_0-T$ denotes the kinetic energy of the electron $T$ measured from the endpoint energy $T_0$.
This type of experimental measurements have been made by Troitsk \cite{Troitsk} and Mainz \cite{Mainz}, and high precision will be achieved by KATRIN \cite{KATRIN}. 

In these experiments the antineutrino escapes undetected; however, magnetic fields select the beta electrons emitted in a particular direction to be studied.
This feature permits the study of anisotropic effects.
In the presence of Lorentz-violating neutrinos, the spectrum \eqref{BetaSpectrum} gets corrected by the replacement \cite{DKL}
\bea\label{Dt_LV}
\De T &\to& \De T + (k_{\mathcal{C}}^{(3)}) + (k_{\mathcal{A}_s}^{(3)})\,\sin\wT \nn\\
&& \qquad\quad + \, (k_{\mathcal{A}_c}^{(3)})\,\cos\wT,
\eea
which shows that the distortion near the endpoint can be shifted and also exhibits a sidereal-time dependence.
The amplitudes in the modification \eqref{Dt_LV} depend of the four independent coefficients $\A{3}{\of}{00}{}$, $\A{3}{\of}{10}{}$, $\A{3}{\of}{11}{}$, $\A{3}{\of}{1-1}{}$ and experimental quantities such as location of the laboratory, orientation of the apparatus, and intensity of the magnetic fields used to select the beta electrons for their analysis \cite{DKL}.

An interesting feature appears when the effective coefficients $\C{2}{\eff}{1m}{}$ are considered, which arise as a consequence of neutrino mass and CPT-even Lorenz violation \cite{KM2012}.
These coefficients can mimic the effects of a mass parameter, in which case the spectrum \eqref{BetaSpectrum} gets modified in the form
\bea
m_\nu^2 &\to& m^2 = m_\nu^2 + (k_{\mathcal{C}}^{(2)}) + (k_{\mathcal{A}_s}^{(2)})\,\sin\wT \nn\\
&& \qquad\quad + \, (k_{\mathcal{A}_c}^{(2)})\,\cos\wT.
\eea
We find that the experimental mass-squared parameter $m^2$ involves the actual neutrino mass $m_\nu$; however, the mass can be screened by the effects of the three coefficients $\C{2}{\eff}{1m}{}$ ($m=0,\pm1$), varying with sidereal time and depending on the orientation of the apparatus.
Notice also that the sign of the experimental mass-squared parameter $m^2$ is not restricted to be positive, so the coefficients $\C{2}{\eff}{1m}{}$ could even mimic a tachyonic neutrino \cite{Chodos}.

Alternative approaches have been considered to search for isotropic Lorentz violation in tritium decay for other operators in Refs. \cite{CarmonaCortes,Bernardini2008}.

\begin{figure}
\centering
\includegraphics[width=0.5\textwidth]{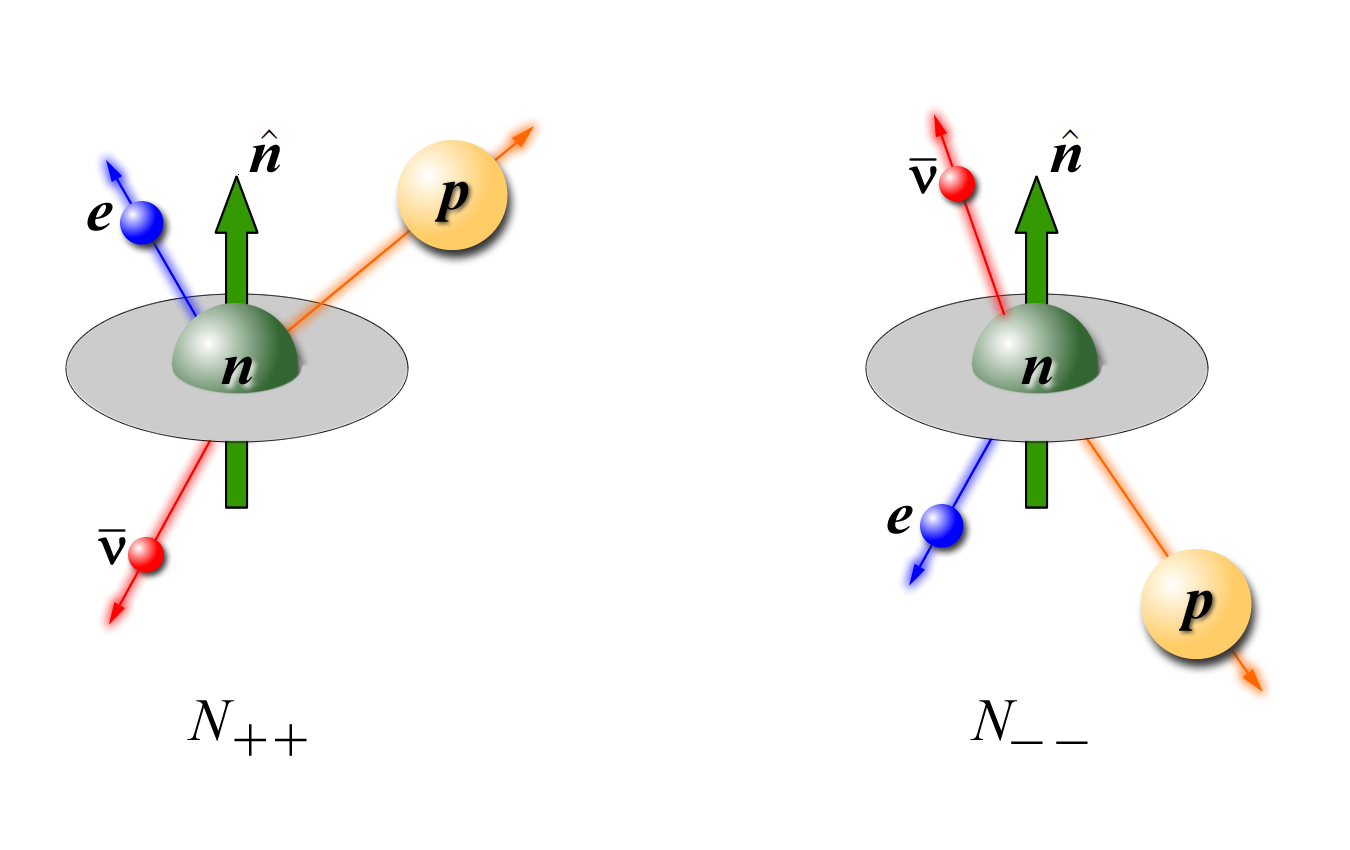}
\caption{Electron-proton coincidence events for the asymmetry $B_\text{exp}$. The polarization of the neutron  is denoted by $\hat{\pmb{n}}$.} 
\label{Figure_B_exp}
\end{figure}

\subsection{Neutron decay}

Neutrons are fascinating laboratories to study the validity of fundamental symmetries.
The effects of deviations from exact Lorentz invariance would affect the spectrum of the beta electrons as well as the measurements of particular experimental asymmetries.
Contrary to tritium decay experiments, the study of neutron decay covers the whole energy spectrum, which takes the form of the spectrum \eqref{BetaSpectrum} neglecting the neutrino mass that plays no role far from the endpoint and the factor $C$ can no longer be approximated by a constant so it becomes a function of the electron energy.
For experiments only counting the number of beta electrons per energy range, all the anisotropic effects disappear after integrating over all the neutrino orientations.
The net effect is a distortion of the whole spectrum that can be studied by searching for deviations from the conventional spectrum.
The residual spectrum is proportional to the coefficient $\A{3}{\of}{00}{}$ \cite{DKL}.

Anisotropic effects can be studied by constructing asymmetries $\pmb{A}_\text{exp}$ in experiments that can determine the directionality of some of the decay products.
For experiments using unpolarized neutrons, an asymmetry counting electrons emitted in the same direction as the antineutrino $N_+$ compared to events in which the two leptons are emitted in opposite directions $N_-$ can be constructed for the measurement of the electron-antineutrino correlation $a$ in the form of 
\beq\label{a_exp}
a_\text{exp} =\frac{N_+-N_-}{N_++N_-}.
\eeq
Similarly, experiments using polarized neutrons that are able to measure the electron and the recoiling proton can be used to search for electron-proton coincidence events, useful for the measurement of the neutrino asymmetry $B$.
As shown in Figure \ref{Figure_B_exp}, the experimental asymmetry counts events in which proton and electron are emitted against $N_{--}$ or along $N_{++}$ the direction of the neutron polarization $\hat{\pmb{n}}$ can be written in the form
\beq\label{B_exp}
B_\text{exp} =\frac{N_{--}-N_{++}}{N_{--}+N_{++}}.
\eeq

Effects of Lorentz violation arise from the modified spinor solutions that affect the matrix element of the decay as well as the unconventional form of the neutrino phase space due to the modified dispersion relations, making the asymmetries \eqref{a_exp} and \eqref{B_exp} have the general form \cite{DKL}
\beq\label{A_exp}
\pmb{A}_\text{exp} = (\pmb{A}_{\mathcal{C}})+ (\pmb{A}_{\mathcal{A}_s})\,\sin\wT + (\pmb{A}_{\mathcal{A}_c})\,\cos\wT,
\eeq
where the amplitudes depend on the coefficients for Lorentz violation $\A{3}{\of}{10}{}$, $\A{3}{\of}{11}{}$, and $\A{3}{\of}{1-1}{}$ and experimental quantities including the orientation and location of the apparatus.

\subsection{Double beta decay}

The same coefficients modifying the spectrum for beta decay can also introduce observable effects in double beta decay experiments.
Since the antineutrinos escape unobserved, the simplest test of Lorentz invariance is an alteration of the two-electron spectrum for the two-neutrino mode of double beta decay produced by the coefficient $\A{3}{\of}{00}{}$ \cite{DKL,Diaz_DBD}.
Similar to the neutron-decay spectrum, the resulting effect is a distortion of the whole spectrum that can be studied by searching for deviations from the conventional spectrum.
The energy at which this effect is maximal has been identified for several isotopes, which will guide these type of studies \cite{Diaz_DBD}.

The neutrinoless mode of double beta decay offers access to other type of coefficient, one that modifies the neutrino propagator.
This Majorana coupling in the SME denoted $|g^{\la\rh}_{\be\be}|$ is a combination of other coefficients in the SME that can trigger neutrinoless double beta decay even if the Majorana mass is negligible.
In terms of this effective coefficient for CPT-odd Lorentz violation, the half-life of an isotope of radius $R$ is given by \cite{Diaz_DBD}
\beq
(T^{0\nu}_{1/2})^{-1} = G^{0\nu}\,|M^{0\nu}|^2\,\frac{|g^{\la\rh}_{\be\be}|^2}{4R^2},
\eeq
where $G^{0\nu}$ is a phase-space factor regarding the two emitted electrons and $M^{0\nu}$ is the relevant nuclear matrix element.
Limits on the Majorana mass parameter $|m_{\be\be}|$ can be used to constrain the coefficient $|g^{\la\rh}_{\be\be}|$.
Since the Lorentz-violating neutrinoless double beta decay depends on the nuclear size $R$ of the isotope used, a future observation of this decay mode can be distinguished because the Majorana-mass mechanism depends on the isotope only through the nuclear matrix elements.

\section{Conclusions}

In this paper we have presented a general overview of the effects of deviations from exact Lorentz invariance in neutrinos in the context of the Standard-Model Extension. 
In general, the signatures of the breakdown of Lorentz symmetry are direction and time dependence of the relevant observables for Earth-based experiments as well as unconventional dependence on the neutrino energy.
Neutrino oscillations are sensitive probes of new physics, which makes this type of experiment an ideal setup to search for violations of Lorentz invariance.
In oscillations, some effects of Lorentz violation include direction and time dependence of the oscillation probability, oscillation phases that grow with the neutrino energy, CPT violation, and mixing between neutrinos and antineutrinos.

Some effects are unobservable in neutrino oscillations, in which case kinematical effects become a complementary technique.
Effects of Lorentz violation appear as modifications to the neutrino velocity as well as unconventional behavior in decay processes. In particular, some decays with neutrinos in the final state can become forbidden above certain threshold energy; similarly, some forbidden processes can become allowed, including \v Cerenkov radiation of one or more particles.
Most of these effects are enhanced by the neutrino energy, which makes high-energy neutrinos of particular interest for future tests of Lorentz invariance.

Finally, there are operators in the theory whose experimental signatures are independent of the neutrino energy. 
In this case, the high precision of low-energy experiments can play a fundamental role in the test of Lorentz symmetry for some particular operators that are unobservable in oscillations and that leave the neutrino velocity unchanged.
For these countershaded operators, beta decay is the ideal experimental setup.
Depending on the properties of the experiment, the main features have been identified for studies of tritium decay, neutron decay, and double beta decay.

To date, there is no compelling evidence of Lorentz violation; nevertheless, only a few of the experimental signatures have been studied \cite{tables}.
Neutrinos offer great sensitivity and numerous ways to test the validity of the cornerstone of modern physics. 
Many of the different techniques presented in this review are currently being implemented by a variety of experimental collaborations.
Interesting new tests of Lorentz symmetry will be performed in the near future, in which low- and high-energy neutrinos will play a key role in our understanding of the nature of spacetime.

\section*{Acknowledgments}
This work was supported in part by the U.S. Department of Energy under grant DE-FG02-13ER42002, the Indiana University Center for Spacetime Symmetries, and the Helmholtz Alliance for Astroparticle Physics (HAP) under Grant no. HA-301.

\end{document}